**High quality epitaxial piezoelectric and ferroelectric wurtzite $Al_{1-x}Sc_xN$ thin films**

*Yang Zeng, Yihan Lei, Yanghe Wang, Mingqiang Cheng, Luocheng Liao, Xuyang Wang, Jinxin Ge, Zhenghao Liu, Wenjie Ming, Chao Li, Shuhong Xie*, Jiangyu Li*, Changjian Li**

Yang Zeng, Shuhong Xie

Key Laboratory of Low Dimensional Materials and Application Technology of Ministry of Education, School of Materials Science and Engineering, Xiangtan University, Xiangtan, Hunan, 411105, China.

E-mail: shxie@xtu.edu.cn

Yihan Lei, Yanghe Wang, Mingqiang Cheng, Luocheng Liao, Jinxin Ge, Zhenghao Liu, Wenjie Ming, Chao Li, Jiangyu Li, Changjian Li

Department of Materials Science and Engineering, Southern University of Science and Technology, Shenzhen, Guangdong, 518055, China.

Guangdong Provincial Key Laboratory of Functional Oxide Materials and Devices, Southern University of Science and Technology, Shenzhen, Guangdong, 518055, China.

E-mail: licj@sustech.edu.cn, lijy@sustech.edu.cn

Shuhong Xie

Key Laboratory of Thin Film Materials and Devices, School of Materials Science and Engineering, Xiangtan University, Xiangtan, Hunan, 411105, China

**Abstract**

Piezoelectric and ferroelectric wurtzite are promising to reshape modern microelectronics because they can be easily integrated with mainstream semiconductor technology. Sc doped AlN ($Al_{1-x}Sc_xN$) has attracted much attention for its enhanced piezoelectric and emerging ferroelectric properties, yet the commonly used sputtering results in polycrystalline $Al_{1-x}Sc_xN$ films with high leakage current. Here we report the pulsed laser deposition of single crystalline epitaxial $Al_{1-x}Sc_xN$ thin films on sapphire and 4H-SiC substrates. Pure wurtzite phase is maintained up to $x = 0.3$ with ≤0.1 at% oxygen contamination. Polarization is estimated to be 140 $\mu C/cm^2$ via atomic scale microscopy imaging and found to be switchable via a scanning probe. The piezoelectric coefficient is found to be 5 times of undoped one when x = 0.3, making it desirable for high frequency radiofrequency (RF) filters and three-dimensional nonvolatile memories.

# 1. Introduction

Wurtzite piezoelectric aluminum nitride (AlN), for its exceptionally thermal conductivity, dielectric breakdown strength, and compatibility with the complementary metal oxide semiconductor (CMOS) process, has a dominating role in bulk acoustic wave (BAW) filters. However, its low piezoelectric coefficient and electromechanical coupling constant limit its applications in future 5G and 6G high bandwidth cellular radios. In 2009, by introducing scandium (Sc) into the AlN lattice, Akiyama et al. reported two-fold enhancement in its piezoelectric coefficient,[1] making it promising for various applications such as high-frequency RF filters,[2-4] high efficiency power generators[5], lead-free high-temperature piezoelectric devices,[2] BAW devices [6-7] and high-electron-mobility transistors (HEMTs).[8-9] More recently, Sc doped AlN has been reported to be ferroelectric with remanent polarization above 100 μC/cm$^2$ and stable up to 600 °C,[10] substantially higher than perovskite Pb(Zr,Ti)$O_3$ or fluoride ferroelectrics, opening new realm for CMOS compatible ferroelectric based technology.[11] Using $Al_{1-x}Sc_xN$, Ferroelectric diode[12] and Ferroelectric field-effect-transistors (FeFET)[13-14] with large memory windows, high on/off ratios, and stability up to 600° C[15] have been demonstrated successfully integrated with Si using a back-end-of-line (BEOL) process, highlighting the potential of $Al_{1-x}Sc_xN$ for scalable, CMOS-compatible, low-power memory and computing applications.

However, current $Al_{1-x}Sc_xN$ films, mostly prepared by magnetron sputtering,[1, 7, 16-18] are polycrystalline in nature, and suffer from large leakage current that degrade their performance. Molecular beam epitaxy (MBE)[19-21] and Metal-organic chemical vapor deposition (MOCVD)[22-24] are also utilized to produce epitaxial $Al_{1-x}Sc_xN$ films, but MBE faces challenges in tuning the phase and stoichiometry[25] of $Al_{1-x}Sc_xN$ while incorporating high Sc contents in MOCVD is difficult due to low volatility of Sc precursor.[22] Hence, improving the quality of epitaxial $Al_{1-x}Sc_xN$ film remains a challenge to realize its full potential for piezoelectric and ferroelectric applications.

Pulsed laser deposition (PLD) is a cheap yet versatile tool for preparation complex oxide thin films, which has been proven effective in preparing high quality epitaxial AlN thin films on various substrates, including sapphire and 4H-SiC.[26-27], however, investigations on pulsed laser deposition (PLD) of epitaxial $Al_{1-x}Sc_xN$ thin films remain limited. Here we report the fabrication of high quality piezoelectric and ferroelectric $Al_{1-x}Sc_xN$ films on sapphire and 4H-SiC single crystal substrates. A phase pure wurtzite structure is maintained up to 30% Sc with minimal oxygen impurity (~ 0.1 at%). The polarization is estimated up to 140 μC/cm$^2$ via atomic scale imaging and switchable via a scanning probe. The piezoelectric coefficient $d_{33}$ increases 5 times with 30% Sc substitution, showing great promises for future high end wave filter and nonvolatile memory applications.

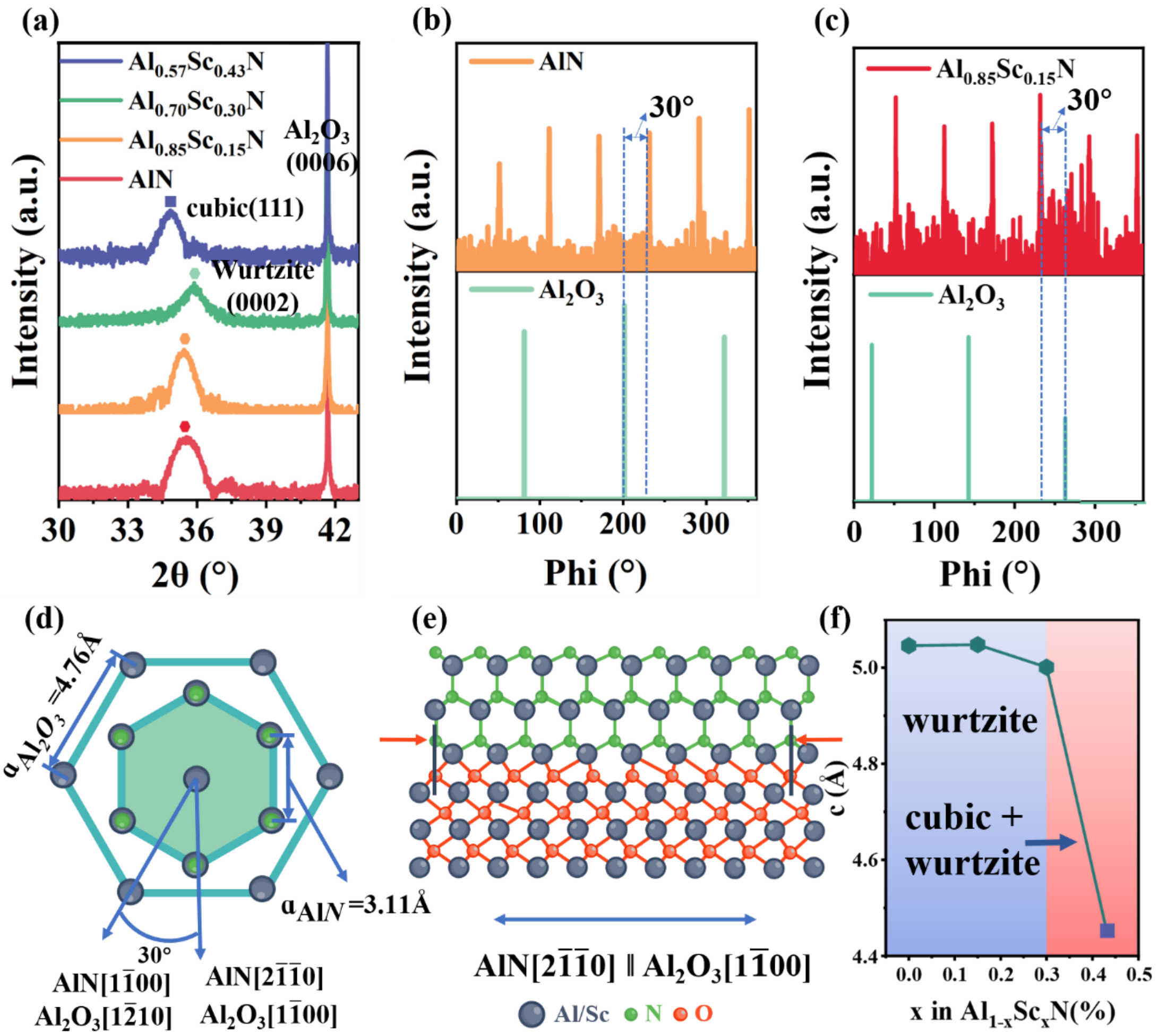

**Figure 1.** Structural characterization of epitaxial $Al_{1-x}Sc_xN$ on single crystalline sapphire substrates. a) The XRD 2$\theta$-$\omega$ scan of $Al_{1-x}Sc_xN$ ($x$ = 0, 0.15, 0.3 and 0.43) films grown on $Al_2O_3$ (0006) substrates. $\varphi$-scan of b) AlN (10$\bar{1}$1) and c) $Al_{1-x}Sc_xN$ (10$\bar{1}$1) grown on $Al_2O_3$ (10$\bar{1}$4). Films peaks (10$\bar{1}$1) clearly indicate a 30° in-plane rotation of the films with respect to the substrate. d) A schematic top view of the epitaxial relationship between $Al_{1-x}Sc_xN$ and $Al_2O_3$. e) A schematic cross-sectional view of the crystallographic model of the $Al_{1-x}Sc_xN/Al_2O_3$ interface, wherein a domain epitaxy relationship is maintained. f) The lattice constant $c$ of films with different Sc components.

## 2. Result

Deposition of high-quality epitaxial $Al_{1-x}Sc_xN$ films are carried using a pulsed laser deposition technique (KrF laser, $\lambda$ = 248 nm) on (0006) $Al_2O_3$ substrates. The laser fluence is 1.6 J/cm$^2$, and the deposition is carried out at substrate temperature of 800 °C under high vacuum conditions (base pressure 3 × $10^{-7}$ torr). The thickness of $Al_{1-x}Sc_xN$ is measured to be ~17 nm. **Figure 1**a shows an X-ray diffraction $2\theta$- $\omega$ scan of deposited films. For Sc concentrations of 0, 0.15, and 0.30, only the (0002) peak of the hexagonal wurtzite phase is observed in $Al_{1-x}Sc_xN$, which sits at ~35.8°. When $x$ = 0.43, another peak is observed at 35.1° that can be attributed to (111) peak of cubic phase $Al_{1-x}Sc_xN$.[28] We thus conclude that we can prepare $Al_{1-x}Sc_xN$ in wurtzite phase up to 0.3 Sc. To investigate the in-plane epitaxial relationship, we performed $\varphi$ scans for the AlN and $Al_{0.85}Sc_{0.15}N$ films (Figure 1b,c). The $Al_2O_3$ substrate shows a 3-fold symmetry while both AlN and $Al_{0.85}Sc_{0.15}N$ peaks show a 6-fold symmetry, and it rotates 30° relative to the substrate peak. Combining the $2\theta$-$\omega$ scan and $\varphi$ scans of the XRD data, we conclude the epaxial relationship as AlN (0001) || $Al_2O_3$(0001) and AlN $[2\bar{1}\bar{1}0]$ || $Al_2O_3$ $[1\bar{1}00]$, which is the same for Sc doped wurtzite structure. Figure 1d is a top view schematic diagram of the epitaxial relationship between $Al_{1-x}Sc_xN$ (0001) and $Al_2O_3$ (0001), which intuitively shows the rotation relationship between the film and substrate. Figure 1e is an interface model between $Al_{1-x}Sc_xN$ and the substrate, which can be described as domain-matching epitaxy (DME) due to the difference in lattice constants,[29-30] where nine times the O-O bond length (9 × 2.747 Å = 24.723 Å) in $Al_2O_3$ nearly equals to eight times the N-N bond length (8 × 3.112 Å = 24.896 Å).[31-32] (the atomic scale images are shown in **Figure S1**, Supplementary Information)Figure 1f shows that the lattice constant $c$ decreases from 5.046 Å to 5.001 Å with increasing Sc content up to 0.3, consistent with the results of previous reports.[1, 33] A sudden decrease in the c-lattice parameter down to 4.45 Å is observed when $x$ = 0.4, corresponding to cubic phase. Hence, the pure wurtzite phase $Al_{1-x}Sc_xN$ is achieved up to $x$ = 0.3 Sc.

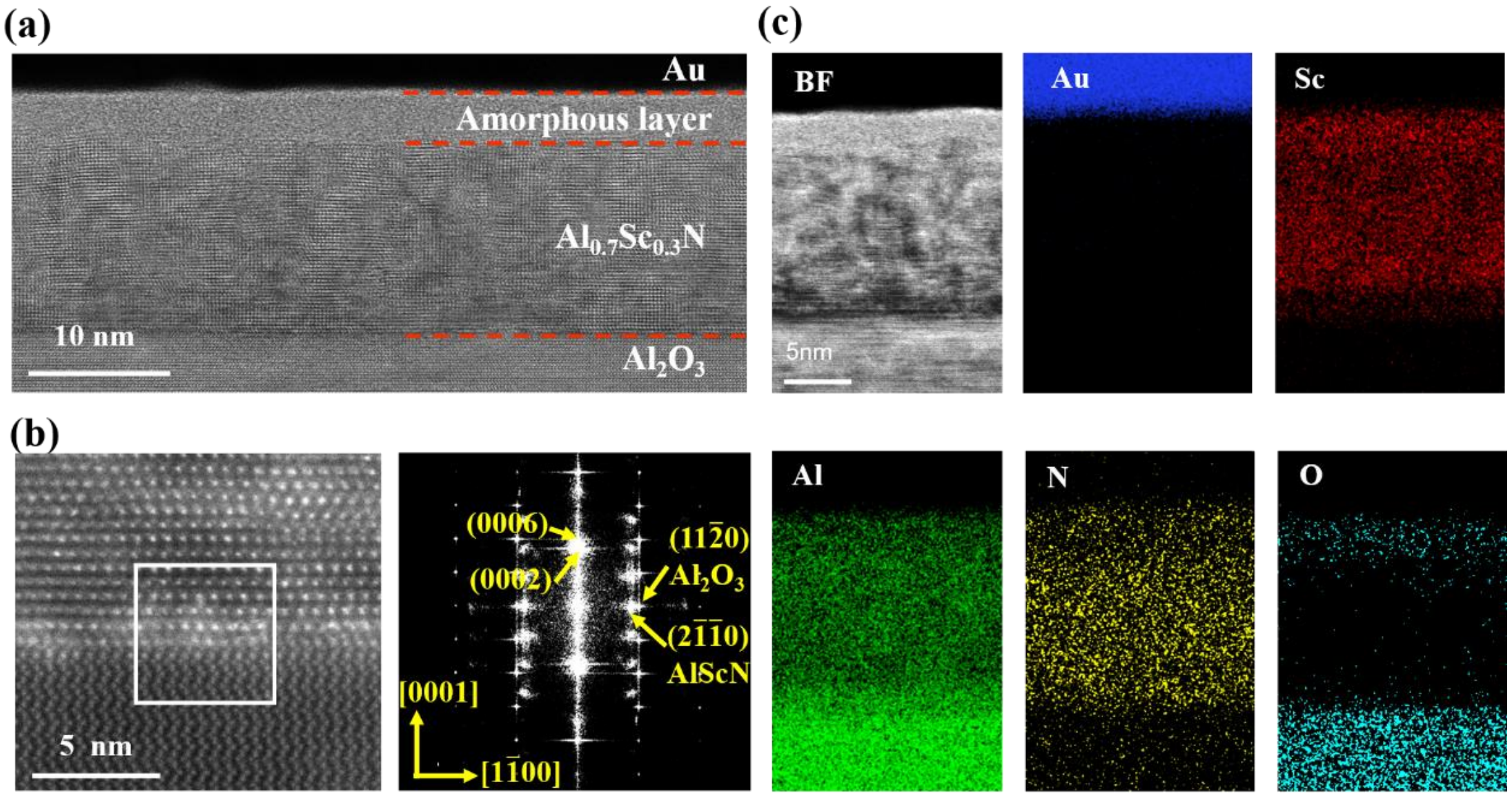

**Figure 2.** Atomic scale characterization of the epitaxial $Al_{0.70}Sc_{0.30}N$ thin films grown on (0006) $Al_2O_3$ substrates. a) BF-STEM cross-section image. b) A typical atomic scale HAADF-STEM image of the $Al_{0.70}Sc_{0.30}N/Al_2O_3$ interface and the corresponding FFT pattern exhibiting the epitaxial relationship. c) Energy-dispersive X-ray spectroscopy (EDS) elemental maps across the cross-section sample.

The quality of the films is further evident from aberration corrected scanning transmission electron microscopy (AC-STEM). A high-resolution bright field-STEM (BF-STEM) image of the $Au/Al_{0.30}Sc_{0.70}N/Al_2O_3$ interface along the $[1\bar{1}00]$ zone axis is shown in **Figure 2**a. $Al_{0.70}Sc_{0.30}N$ layer has a uniform thickness of ~17 nm, and a ~3.5 nm surface amorphous layer is present, which is most likely caused by surface oxidation, evident from energy dispersive spectroscopy (EDS) mapping. A zoomed-in high angle annular dark field (HAADF) STEM image of the $Al_{0.70}Sc_{0.30}N/Al_2O_3$ is shown in Figure 2b, where an atomic sharp interface is observed. From the fast Fourier transformation (FFT) pattern across the interface (white square enclosed region), a good epitaxy relationship of $Al_{0.70}Sc_{0.30}N$ (0001) || $Al_2O_3$ (0001) and $Al_{0.70}Sc_{0.30}N$ $[2\bar{1}\bar{1}0]$ || $Al_2O_3$ $[11\bar{2}0]$ is confirmed, consistent with previous XRD results. As shown in Figure 2c, EDS elemental mappings confirm that Al, Sc and N distribute uniformly in deposited $Al_{0.70}Sc_{0.30}N$ film. Surface oxygen layer is likely resulted by the oxidation after deposition. Since signal noise ratios of N and O K-edge quantification in EDS measurements are not high enough, we further performed EELS studies, which shows no oxygen peak (**Figure S2**, Supplementary Information) in the interior of the $Al_{0.70}Sc_{0.30}N$ films. The estimated maximum of oxygen contamination is ~ 0.1 at% using typical EELS collection conditions [34,35]. The combined EDS and EELS studies confirm that our PLD has good crystallinity and composition control in $Al_{0.70}Sc_{0.30}N$ deposition process. The quality of film is evident from ultralow leakage current density $5\times10^{-5}$ $A/cm^2$ at an electric field of 4 MV/cm (**Figure** S3, Supplementary Information), which is lowest among the existing reports (Table S1, Supplementary Information).

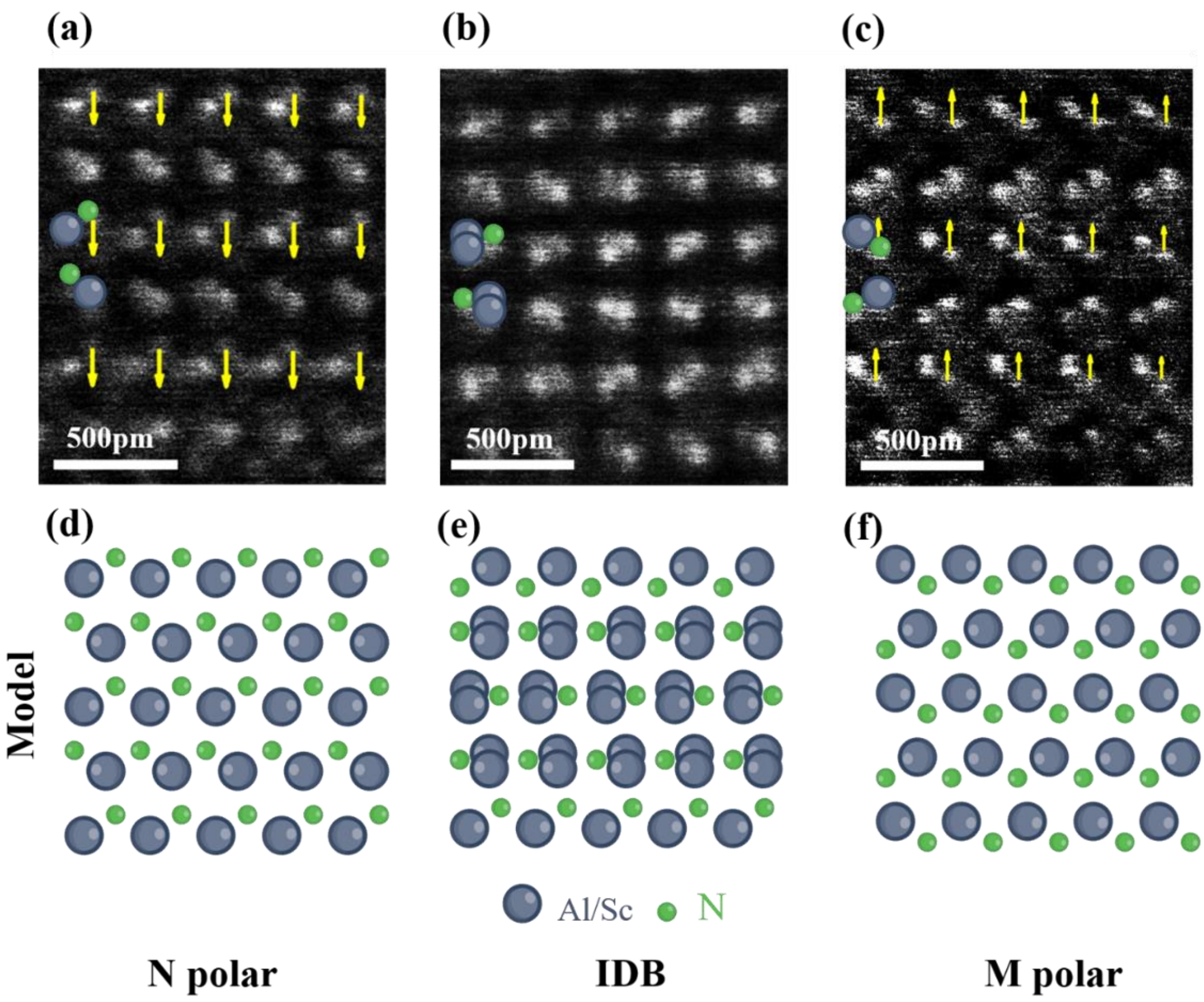

**Figure 3.** Atomic characterization of the polarization via contrast inverted ABF-STEM images. a-c) Inverted ABF-STEM images of the $Al_{0.70}Sc_{0.30}N$. From left to right is N polar, inversion domain boundaries (IDB) region, and M polar. Green and dark blue balls represent the N and Al/Sc atoms. The uc-by-uc c-axis atomic displacement between Al/Sc and N is superimposed on corresponding image, where the length of the arrows indicates the magnitude of polarization. d-f) N-polar, IDB and M-polar atomic models of $Al_{0.70}Sc_{0.30}N$.

Furthermore, we investigated the polarity of $Al_{0.70}Sc_{0.30}N$ using inverted ABF-STEM. We identify metal M(Al/Sc)-polar and N-polar domains as well as the inversion domain boundary (IDBs)[32, 36-40] in between in the as-deposited $Al_{0.70}Sc_{0.30}N$ films (**Figure 3**a). As shown in the corresponding atomic schematics (Figure 3b), N atoms sit above the Al/Sc atoms in the N polar region, with a polarization pointing downwards. In M polar region, N atoms sit below the metal atom, so that the polarization points upwards (Figure 3d). The transition region is IDB where N atoms almost sit at the same z position as metal atoms (Figure 3c).

Using a center-of-mass peak fitting method,[41] we could identify the N and Al/Sc positions accurately (as small as 5 pm), so that we could determine the relative displacement of metal cations in the out-of-plane direction with respect to N anions as 70 pm in both N-polar and and M-polar regions. Using the Born effective charge ($Z^B = 2.5e$), we can evaluate the spontaneous polarization as,[10, 37, 42]

$$P=\frac{1}{V}Z^{B}\Delta d \quad (1)$$

where V is the volume of the unit cell ( $V=\frac{\sqrt{3}}{2}a^2c$) and $\Delta d$ is the N to M displacement in $c$ direction (yellow arrows). The polarization is found to be 140 μC/cm$^2$, comparable to theoretical predications [43] and previous experimental results. [10, 37]

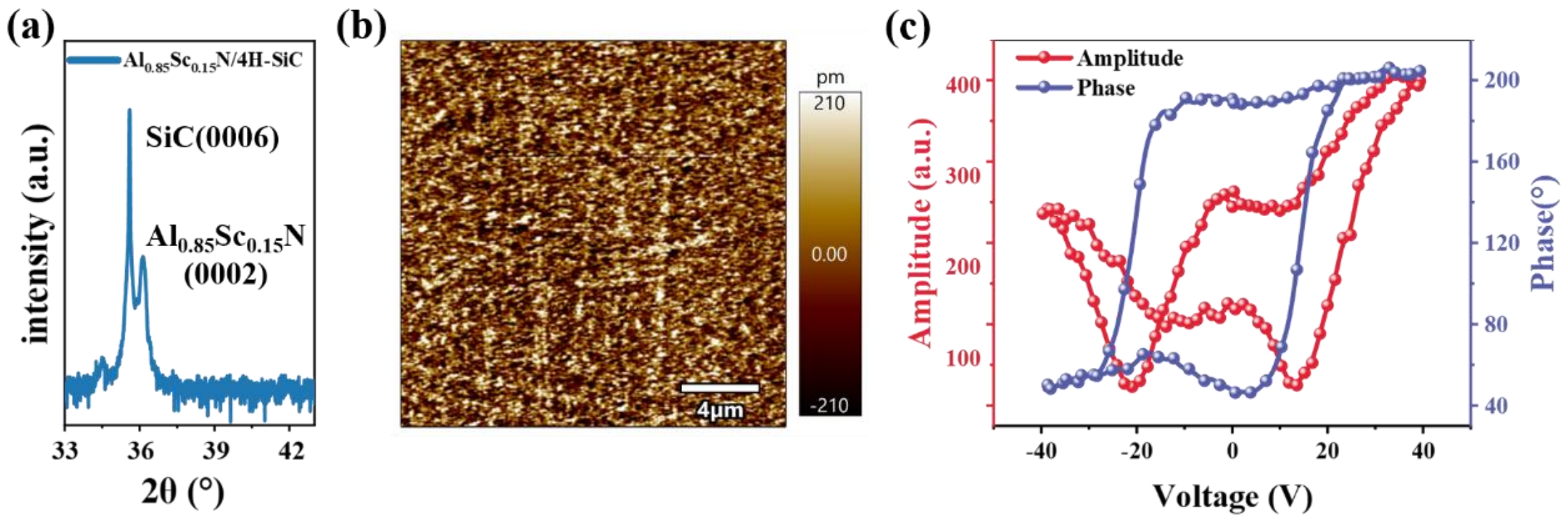

**Figure 4.** a) The XRD 2$\theta$-$\omega$ scan of $Al_{1-x}Sc_xN$ grown on 4H-SiC. b) The AFM morphology image of the thin film, which exhibits a relatively smooth surface. c) The PFM single-point test on the sample on 4H-SiC.

In order to characterize the ferroelectric properties of $Al_{1-x}Sc_xN$, we also prepare epitaxial films on conductive 4H-SiC substrates. As shown in **Figure 4**a, a single strong peak adjacent to the substrate (0006) peak in the XRD spectrum indicates that a good epitaxy is maintained. A typical AFM image (Figure 4b) of the $Al_{0.85}Sc_{0.15}N$ grown on 4H-SiC substrate also shows a small surface roughness ~200 pm. Piezoresponse force microscopy (PFM) amplitude versus voltage curve shows a characteristic butterfly loop for ferroelectric materials (Figure 4c), while the accompanying phase-voltage hysteresis loop shows a phase difference close to ~180°, confirming the ferroelectric switching. Coercive voltage is found to be +13 V and -20 V, respectively, and the imprint is likely caused by the asymmetric electrodes.

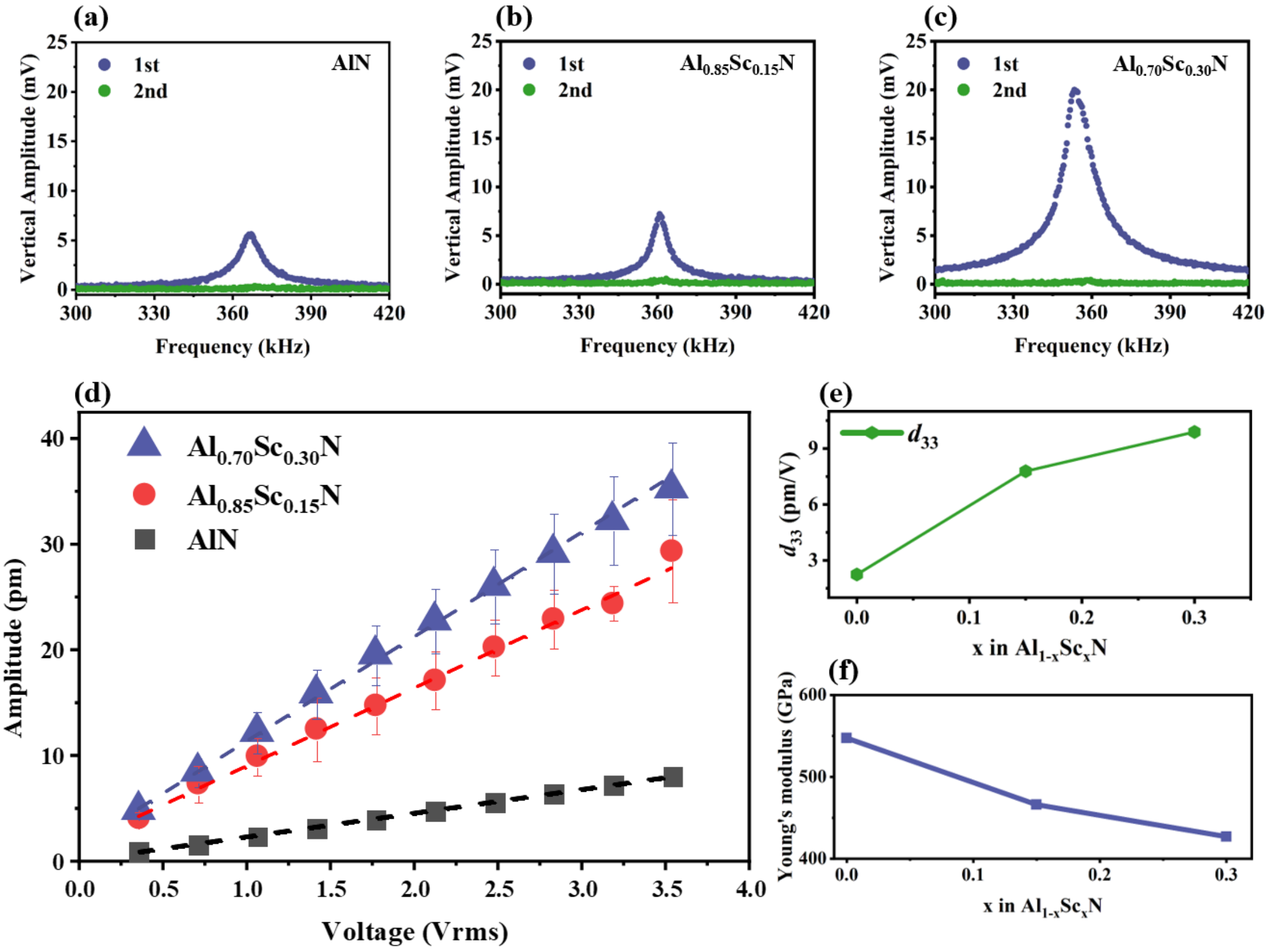

**Figure 5.** Piezoelectric properties measurement of $Al_{1-x}Sc_xN$ films via PFM. a-c) Vertical piezoresponse amplitude versus excitation frequency for a) AlN, b) $Al_{0.85}Sc_{0.15}N$ and c) $Al_{0.70}Sc_{0.30}N$ films. d) The vertical amplitude of $Al_{1-x}Sc_xN$ films exhibits a linear relationship with AC voltage for $x$=0, 0.15, and 0.30. e) Based on the results obtained from (d), the $d_{33}$ values of the films at different concentrations can be determined. f) young's modulus of the $Al_{1-x}Sc_xN$ films with varying Sc concentration.

Furthermore, the epitaxial $Al_{1-x}Sc_xN$ ($x$ = 0, 0.15, 0.30) films show enhanced piezoelectric performance with increasing Sc concentration as revealed by PFM measurements. To exclude other extrinsic factors, both the first and second harmonic piezoresponses are tested following our previous methods.[44-45] The vertical amplitude response, as shown in **Figures 5**a-c, demonstrate that the first harmonic response dominates the second harmonic one under a 2V AC driving voltage, confirming the intrinsic piezoelectric response in wurtzite $Al_{1-x}Sc_xN$ films when $x \leq 0.3$. In contrast, the second harmonic response is only slightly lower than the first harmonic response in $Al_{0.6}Sc_{0.4}N$, exhibiting non-piezoelectric behavior (Figure S4 in Supplementary Information).

From Figure 5a-c, we also see the resonant peak shifts to lower frequency with increased Sc, suggesting the softening of the film. The effective piezoelectric coefficient $d_{33}$ can be estimated from PFM by measuring the amplitude responses under varying AC drive voltages, and a good linear relationship is observed in $Al_{1-x}Sc_xN$ ($x$ = 0, 0.15, and 0.30) films (Figure 5d). As shown in Figure 5e, the effective $d_{33}$ increases from 2 pm/V to 10 pm/V when x reaches to 0.3, which is about five times of the undoped AlN. It is worth noting that PFM measurements underestimate [46] the piezoelectric coefficient compared to bulk measurements, yet relative comparison is clear, and we expect much larger intrinsic $d_{33}$ for our $Al_{0.70}Sc_{0.30}N$ film.

We also carry out the Young's Modulus measurement on the $Al_{1-x}Sc_xN$ as it is important for electromechanical coupling coefficient ($K^2 = {d_{33}^2 E_3}/{\varepsilon_r \varepsilon_0}$, where $E_3$ is the Young's modulus, $\varepsilon_r$ and $\varepsilon_0$are the relative and vacuum permittivity, respectively.) when used for RF filter applications.[47] The left shift resonant frequency with increasing Sc indicates a softer film,[48] and it is confirmed by the Young's modulus data measured by the AM/FM mode of AFM.[49] As shown in Figure 5f, the Young's modulus data decreases from 547 GPa to 427 GPa when x increases from 0 to 0.3 (histograms shown in Figure S5 in Supplementary Information), consistent with previous experimental reports.[50-51] While $E_3$ and $d_{33}$ show opposite trends, it is expected that $K^2$ increases with increasing Sc concentration as the enhancement in $d_{33}$ is much stronger.

## 3. Summary

In summary, we successfully fabricated high-quality epitaxial $Al_{1-x}Sc_xN$ thin films on $Al_2O_3$ and 4H-SiC substrates using PLD technique. Pure wurtzite phase is maintained when $x \leq 0.3$. Atomic resolution HAADF-STEM images reveal superior crystal quality, with polarization values ~140 μC/cm$^2$, which is switchable by scanning probe. Furthermore, the piezoelectric coefficient of Sc doped AlN films is enhanced 5 times compared to undoped one when $x = 0.3$. The enhanced piezoelectric and ferroelectric properties of our $Al_{1-x}Sc_xN$ films make them promising for a wide range of electromechanical and memory applications.

## 4. Experimental Section

**Sample Fabrication.** $Al_{1-x}Sc_xN$ thin films were prepared using a pulsed laser deposition technique. Sintered AlN, $Al_{0.8}Sc_{0.2}N$ and $Al_{0.6}Sc_{0.4}N$ ceramic targets are used for deposition. The deposition process was carried out on (0006) $Al_2O_3$ single crystal substrates at a temperature of 800 ° C and a pressure of $3\times10^{-7}$ torr. A laser fluence of 1.6 J/mm$^2$ is maintained for all depositions. After deposition, samples were cooled under the same vacuum conditions at a rate of 20 °C /min to room temperature. $Al_{1-}$

$_x Sc_x N$ epitaxial thin films ranged from 5-100 nm thickness were prepared. The calibrated growth rate is ~1.7Å/s. The Al/Sc ratio of the films are calculated from EDS measurements.

**Characterization**. XRD analysis was performed using a Rigaku Smartlab XRD system. STEM imaging was conducted using a FEI Titan Themis G2 electron microscope from Thermo Fisher Scientific, operated at 300 kV. For HAADF-STEM image, the probe convergence semi-angle is 25 mrad and collection inner and outer semi-angles are 45 mrad and 200 mrad, respectively. PFM is performed using MFP-3D Infinity from Oxford instruments, using the ASYELEC-01-R2 tip.

**Acknowledgements**

Y. Zeng and Y. Lei contributed equally to this work. This work was supported by National Natural Science Foundation of China (Grant Nos. 52172115), Nature Science Foundation of Guangdong Province, China (Grant Nos. 2022A1515010762) and supported by Shenzhen Science and Technology Program (Grant Nos. 20231121093057002), and Outstanding Talents Training Fund in Shenzhen.

*Supporting Information* for

**High quality epitaxial piezoelectric and ferroelectric wurtzite $Al_{1-x}Sc_xN$ thin films**

*Yang Zeng, Yihan Lei, Yanghe Wang, Mingqiang Cheng, Luocheng Liao, Xuyang Wang, Jinxin Ge, Zhenghao Liu, Wenjie Ming, Chao Li, Shuhong Xie*, Jiangyu Li*, Changjian Li**

Yang Zeng, Shuhong Xie

Key Laboratory of Low Dimensional Materials and Application Technology of Ministry of Education, School of Materials Science and Engineering, Xiangtan University, Xiangtan, Hunan, 411105, China.

E-mail: shxie@xtu.edu.cn

Yihan Lei, Yanghe Wang, Mingqiang Cheng, Luocheng Liao, Jinxin Ge, Zhenghao Liu, Wenjie Ming, Chao Li, Jiangyu Li, Changjian Li

Department of Materials Science and Engineering, Southern University of Science and Technology, Shenzhen, Guangdong, 518055, China.

Guangdong Provincial Key Laboratory of Functional Oxide Materials and Devices, Southern University of Science and Technology, Shenzhen, Guangdong, 518055, China.

E-mail: licj@sustech.edu.cn, lijy@sustech.edu.cn

Shuhong Xie

Key Laboratory of Thin Film Materials and Devices, School of Materials Science and Engineering, Xiangtan University, Xiangtan, Hunan, 411105, China

**1. ABF-STEM image of domain-matching epitaxy between $Al_{0.70}Sc_{0.30}N$**

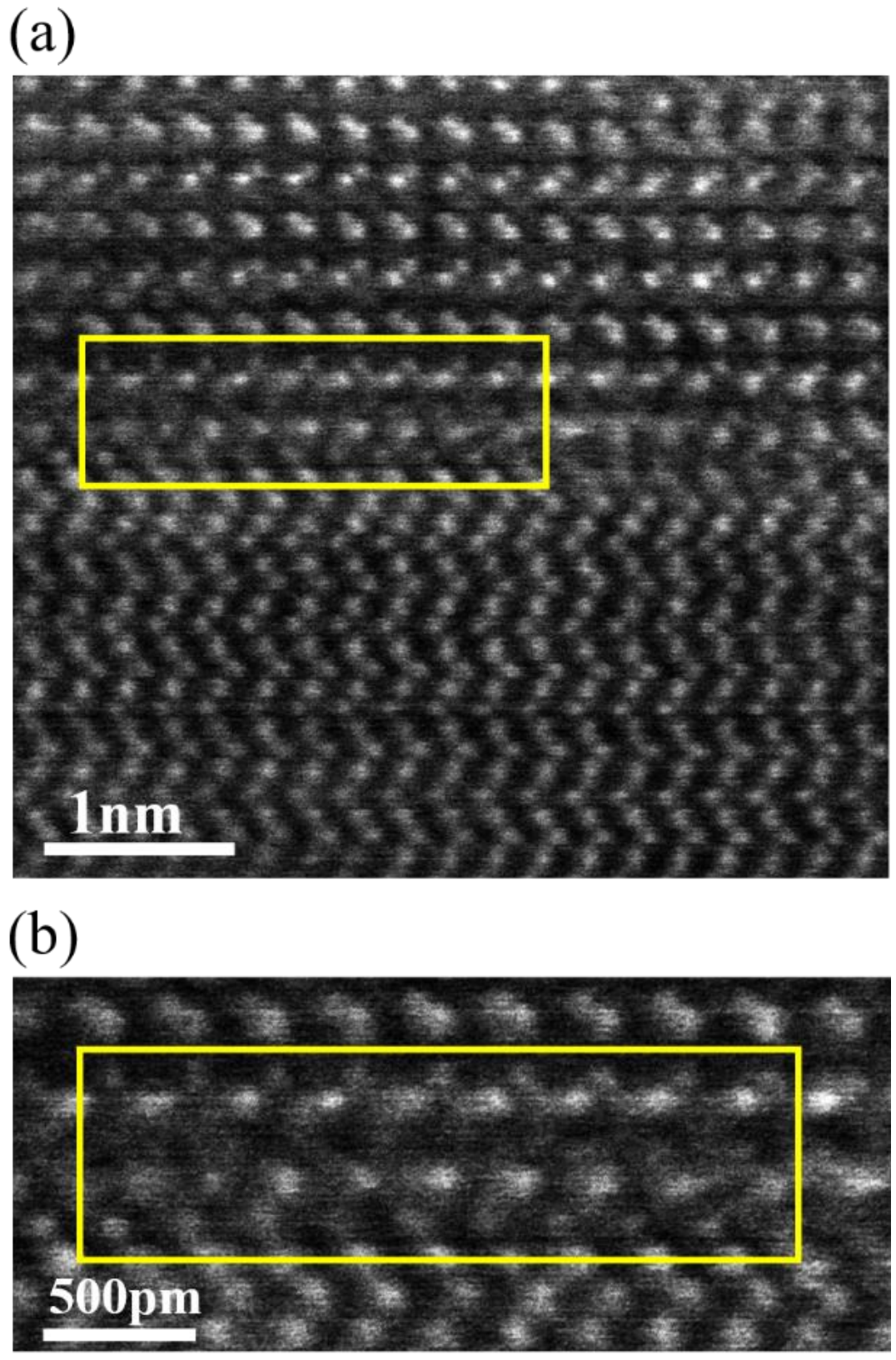

**Figure S1**. a) A contrast inverted ABF-STEM image of the cross-sectional $Al_{0.70}Sc_{0.30}N/Al_2O_3$ interface. b) The zoomed ABF-STEM image from the yellow rectangle enclosed region from Figure (a) showing a 9 × O-O bond length and 8 × N-N bond length.

A contrast inverted ABF-STEM across the $Al_{0.70}Sc_{0.30}N/Al_2O_3$ interface shows an sharp interface. It can be observed that N-polarity AlN is present on top of the $Al_2O_3$ substrate. By zooming in on the region, Figure S1b reveals 9 × O-O bonds corresponding to 8 × N-N bonds, which aligns with the original Figure 1e in the text.

**2. EELS spectroscopy of the epitaxial $Al_{0.70}Sc_{0.30}N$ thin films**

To characterize the possible existence of oxygen contamination, we performed STEM-EELS on epitaxial $Al_{0.70}Sc_{0.30}N$ thin films as EELS is more sensitive to detect oxygen compared to EDS. **Figure S2**a shows the HAADF-STEM images showing the EELS spectra acquired from the yellow box enclosed region, and N and O intensity maps are shown, respectively. From the N and O EELS elemental mapping results, we identify that O is present near the surface (spectra 1) and interface region (spectra 3). The region 1 corresponds to the surface oxidation effect and the interfacial oxygen is probably due to diffusion of oxygen from the $Al_2O_3$. The O *K* edge is missing in spectra 2, indicating the oxygen contamination is minimal. Under typical conditions, the oxygen atom detection limit is ~ 0.1 at%, which sets the upper limit for oxygen contamination.

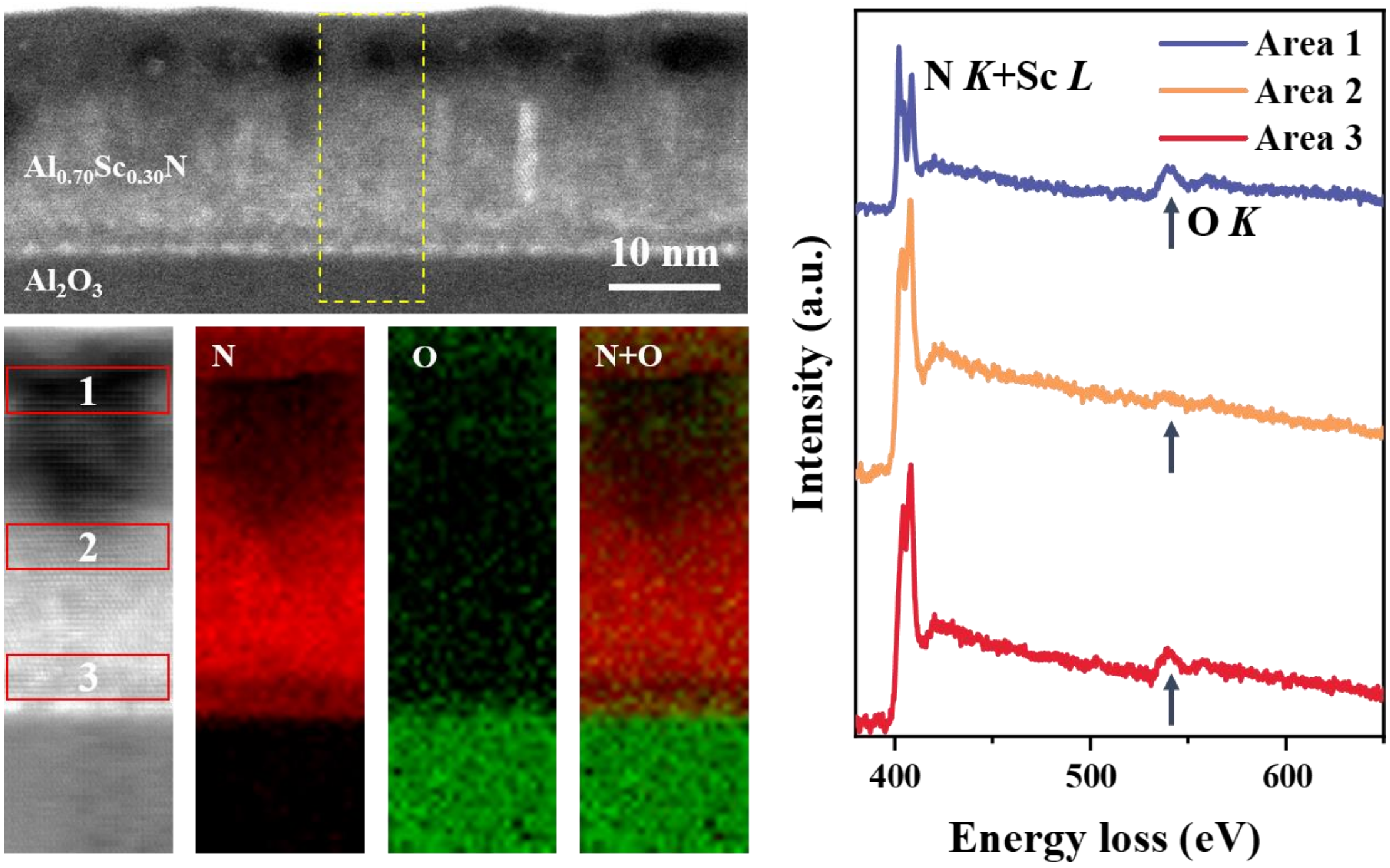

**Figure S2**. STEM-EELS cross-sectional image of the $Al_{0.70}Sc_{0.30}N/Al_2O_3$ interface. a) A HAADF image of $Al_{0.70}Sc_{0.30}N/Al_2O_3$ heterostructure. b) An ADF-STEM image and N, O elemental mapping results acquired simultaneously at the yellow rectangle enclosed area in (a). c) The STEM-EELS electron energy loss spectroscopy from the surface, interior and surface region of $Al_{0.70}Sc_{0.30}N$ films.

## 3. Leakage current of epitaxial $Al_{0.70}Sc_{0.30}N$ thin films

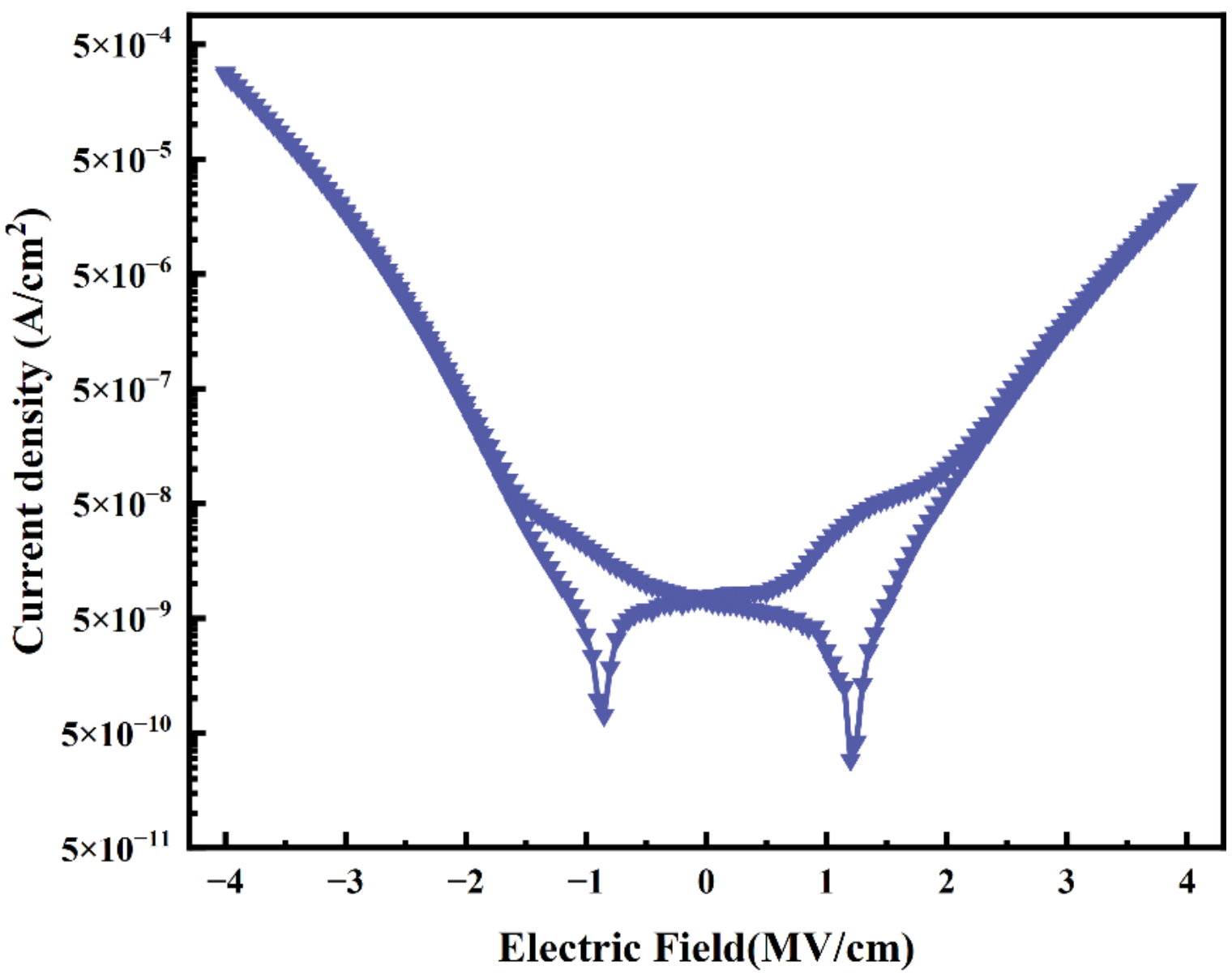

**Figure S3.** DC measurement of leakage current of PLD prepared 17 nm Al0.70Sc0.30N thin films with a Φ=100 μm top electrode.

## 4. The electromechanical response of $Al_{0.57}Sc_{0.43}N$ films

PFM characterization of $Al_{0.54}Sc_{0.47}N$ film is performed using first and second harmonic response. The piezoelectric response is mainly manifested in the first harmonic response, while the electrostrictive response is primarily reflected in the second harmonic. When the piezoelectric response is very weak or absent, the intensity of the second harmonic response will dominate. From **Figure S4**, it can be observed that the second harmonic response is only slightly lower than the first harmonic response, indicating that the sample exhibits non-piezoelectric behavior in this case.

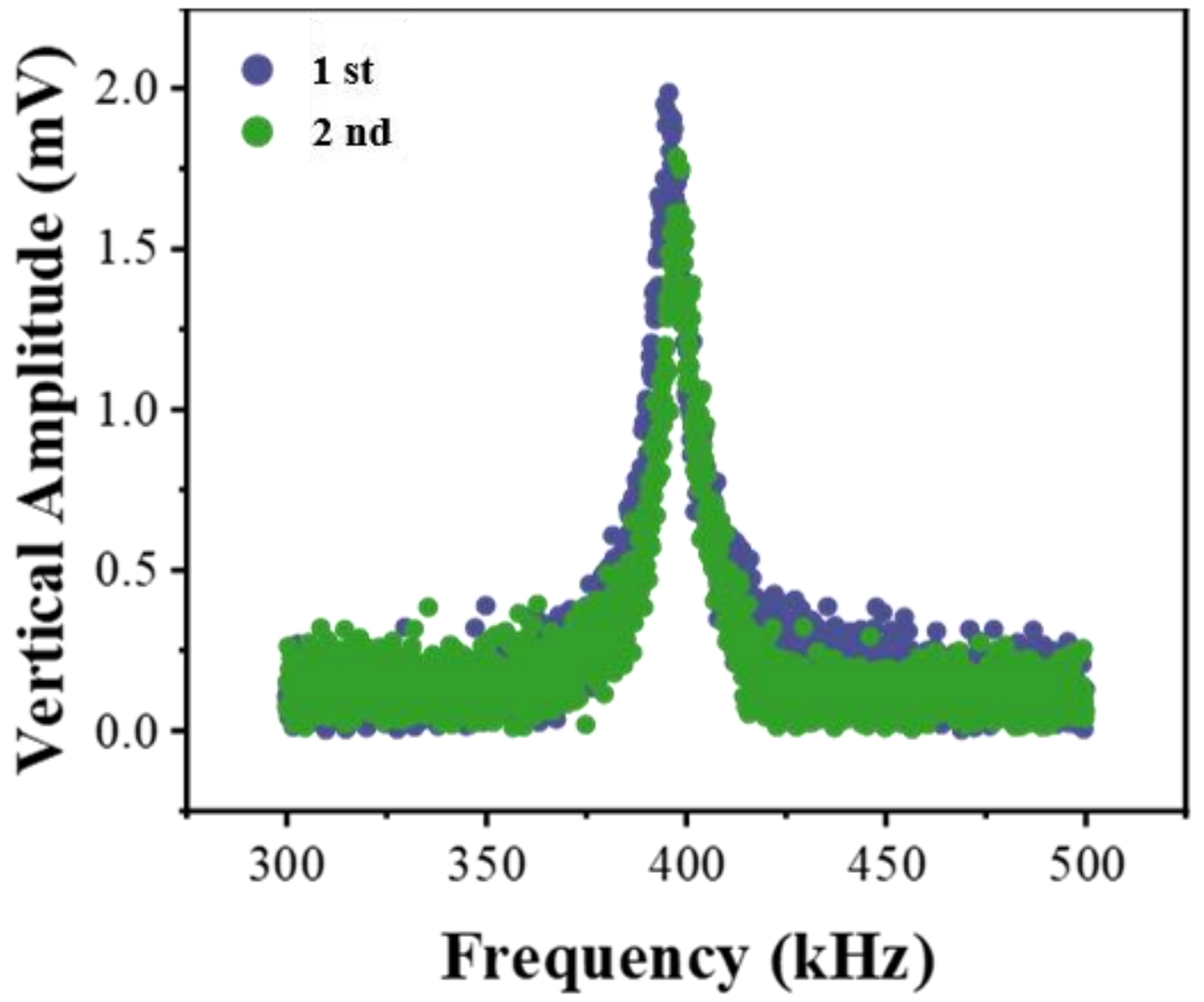

**Figure S4**. The first and second harmonic responses in the vertical direction of $Al_{0.57}Sc_{0.43}N$ during PFM measurements.

**5. Young's modulus of $Al_{1-x}Sc_xN$**

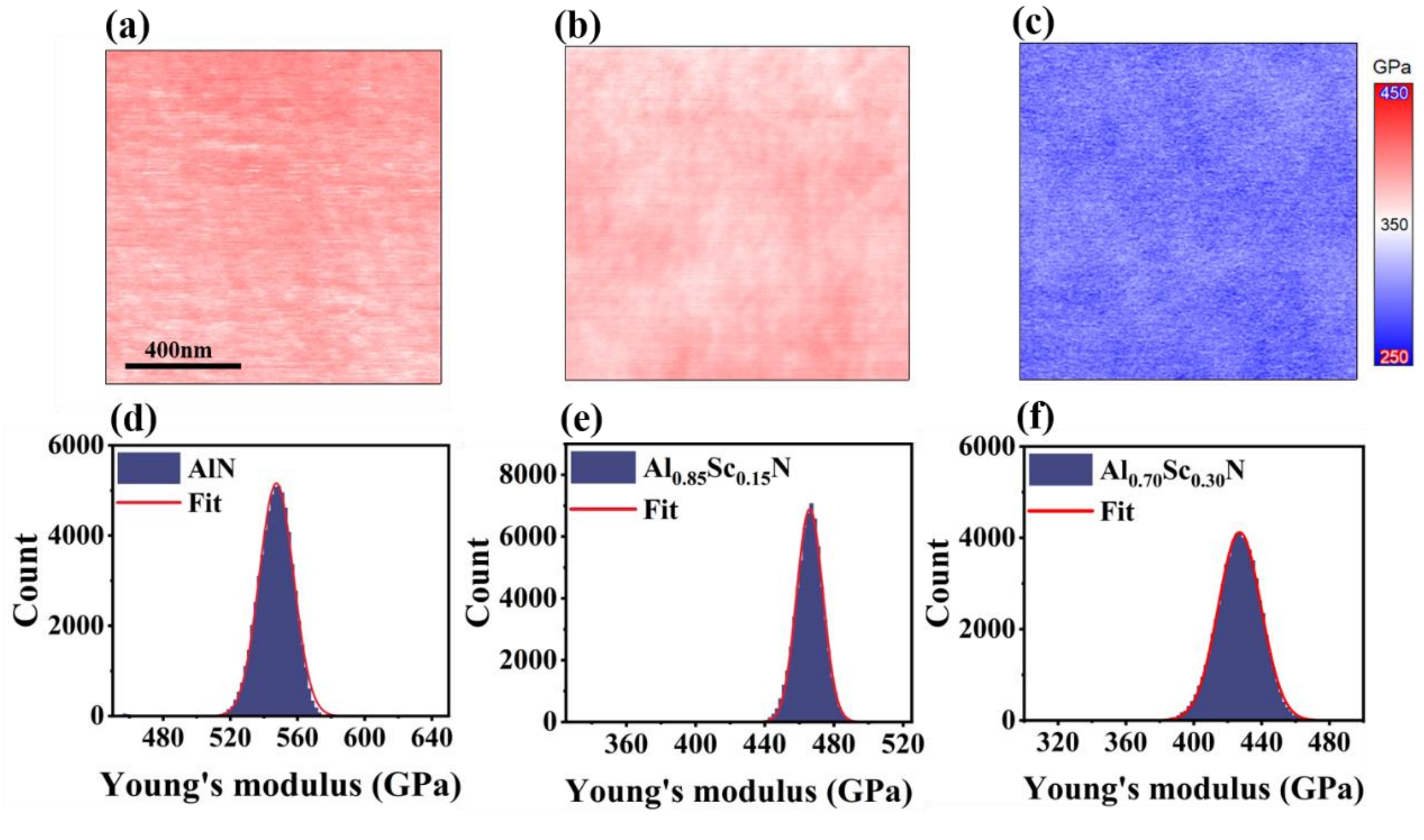

**Figure S5**. a-c) The mapping of Young's modulus for $Al_{1-x}Sc_xN$ films. a) AlN. b) $Al_{0.70}Sc_{0.30}N$. c) $Al_{0.85}Sc_{0.15}N$. d-f) The histogram distribution of Young's modulus for $Al_{1-x}Sc_xN$ films. d) AlN. e) $Al_{0.70}Sc_{0.30}N$. f) $Al_{0.85}Sc_{0.15}N$.

**Table S1** Comparison of materials properties of $Al_{1-x}Sc_xN$ grown by various methods.

| Growth technique | Substrate /crystallinity | Sc content | Growth Temperature (℃) | Thickness (nm) | Leakage Current (A/ $Acm^{-2}$) | Pr ($\mu C/cm^2$) |
|---|---|---|---|---|---|---|
| Sputter | Si/ Texture[1] | 0-0.45 | 500 | ~1100 | \ (no data) | \ |
| | Si/ Texture[2] | 0.22 | 400 RT | 50 | ~130 $A/cm^2$ 165 $A/cm^2$ (PE) | 115 65 |
| | Si/ Texture[3] | 0.25-0.35 | 300 | 40 | ~70 $A/cm^2$ (PE) | 85-150 |
| | Si/ Texture[4] | 0.29 | 350 | 100 | ~$10^{-8}$ A (DC) | 140 |
| | Silica glass/ Texture[5] | ~0.6 | \ | 4000-5000 | \ | \ |
| | Sapphire/ Texture[6] | 0.28 | 450 | 100/200/300 | ~22 $A/cm^2$ (PE) | 150/135/125 |
| | Sapphire/ Texture[7] | 0.25 | 450 | 550 | \ \ | \ |
| MBE | Sapphire/ Single crystalline[8] | 0.3 | 700-900 | 27 | ~30 $A/cm^2$ (PE) | 150 |
| | Sapphire/ Single crystalline [9] | 0.14-0.36 | 700-900 | 100 | 25~300 $A/cm^2$ | 115-155 |
| | SiC, GaN/ Single crystalline [10] | 0.16-0.20 | 360-890 | 80 | \ | \ |
| | Sapphire/ Single crystalline [11] | 0.21 | 700-900 | 100 | ~3$A/cm^2$ (PE) | \ |
| | Sapphire/ Single crystalline [12] | 0.15 | 1000 | 230 | ~30 | 165 |
| MOCVD | Sapphire/ Single crystalline [13] | 10.9-15.8 | 900-1200 | 10 | \ | \ |
| | Sapphire/ Single crystal[14] | ~30 | 1000 | 10-100 | \ | \ |
| **PLD (this work)** | Sapphire，SiC/ Single crystalline | 0-0.43 | 800 | 17 | $3\times10^{-9}$A /$5\times10^{-5}$ $A/cm^2$ (DC) | 140 (STEM) |

## Refereneces